# Hints of the Quantum Nature of the Universe in Classical Electrodynamics and Their Connection to the Electronic Charge and Dark Energy

**Vernon Cooray [1], Gerald Cooray [2], Marcos Rubinstein [3] and Farhad Rachidi [4]**

[1] Department of Electrical Engineering, Uppsala University, 752 37 Uppsala, Sweden.
[2] Karolinska Institute, Stockholm, Sweden
[3] HEIG-VD, University of Applied Sciences and Arts Western Switzerland, 1401 Yverdon-les-Bains, Switzerland
[4] Electromagnetic Compatibility Laboratory, Swiss Federal Institute of Technology (EPFL), 1015 Lausanne, Switzerland

*Abstract*: The electromagnetic fields of linear radiating systems working without dispersive and dissipative losses are analyzed both in the time and the frequency domains. In the case of the time domain radiating system, the parameter studied is the action, *A*, associated with the radiation. The action is defined as the product of the energy and the duration of the radiation. In the case of the frequency domain radiating system, which produces radiation in bursts of duration *T/2* where *T* is the period of oscillation, the parameter studied is the energy, *U*, dissipated in a single burst of radiation of duration *T/2*. In this paper, we have studied how *A* and *U* vary as a function of the charge associated with the current in the radiating system and the ratio of the length of the radiating system and its radius. We have observed remarkable results when this ratio is equal to the ratio of the radius of the universe to the Bohr radius. In the case of the time domain radiating system, we have observed that when the charge associated with the current propagating along the radiator reaches the electronic charge, the action associated with the radiation reduces to $h/2\pi$ where *h* is the Planck constant. In the case of the frequency domain radiating system, we have observed that as the magnitude of the oscillating charge reduces to the electronic charge, the energy dissipated in a single burst of radiation reduces to $h\nu$, where $\nu$ is the frequency of oscillation. Interestingly, all these results are based purely on classical electrodynamics and general relativity. The importance of the findings is discussed. In particular, the fact that the minimum free charge that exists in nature is the electronic charge, is shown for the first time to be a direct consequence of the photonic nature of the electromagnetic fields. Furthermore, the presented findings allow to derive for the first time an expression for the dark energy density of the universe in terms of the other fundamental constants in nature, the prediction of which is consistent with experimental observations. This Equation, which combines together the dark energy, electronic charge and mass, speed of light, gravitational constant and Planck constant, creates a link between classical field theories (i.e., classical electrodynamics and general relativity) and quantum mechanics.

## 1. Introduction

Classical electrodynamics is an old subject which has its origins in the 19[th] century with the pioneering work due to James Clerk Maxwell. The subject has been developed and expanded extensively over the years and as it stands today, it is a subject which is fully matured and thoroughly explored. However, as we will show in this paper, there are interesting features hidden within classical electrodynamics which went unnoticed for nearly 160 years by scientists who were more interested either in the basic structure of electrodynamics or its applications in practice.

In analyzing the electromagnetic fields generated by long transmitting antennas including lightning flashes, scientists and engineers utilize electromagnetic field equations pertinent to various charge and current distributions both in time and frequency domain. The expressions for the radiation fields generated by long transmitting antennas, the subject matter of this paper, are familiar to the scientists and engineers working in the field of antenna theory and lightning research. This indeed is the case because lightning flashes are perhaps the longest electromagnetic radiators that exist on Earth. Interestingly, the standard and familiar equations that are being used frequently by scientists and engineers to obtain electromagnetic radiation fields of long antennas including lightning flashes contain hidden features that were only recently discovered [1, 2, 3]. The goal of this paper is to improve some of the calculations presented in the above referenced papers and to illustrate these hidden features of the electromagnetic radiation fields.

## 2. Time domain electromagnetic fields

### 2.1 Time domain radiation fields generated by a long antenna

Consider a long and vertical antenna located over a perfectly conducting ground plane. The perfectly conducting ground plane coincides with the plane x-y with z = 0. The relevant geometry is shown in Figure 1. A current is injected into this antenna from the ground end (z = 0). Recall that we are considering the ideal case without any losses. That is, we neglect all the losses such as thermal, dispersive and dissipative. In this case, the current pulse propagates along the antenna without attenuation and dispersion. This is exactly identical to the transmission line model, which is one of the models used to simulate the electromagnetic fields generated by lightning return strokes [4]. We consider the case in which the current pulse gets absorbed when it reaches the end of the antenna, i.e., there is no current reflection. Let us represent the current waveform that is being injected at the base of this antenna by $i(t)$. The current pulse propagates upwards with speed $v$. Here, we are only interested in the radiation fields and we consider a point of observation at a distance $r$ which is much larger than the length of the antenna $L$. The effect of the perfectly conducting ground plane is taken into account by using image theory.

The radiation field generated by the antenna consists of four radiation field pulses. The first pulse is created during the initiation of the current pulse at the base of the antenna. The second one is created by the initiation of the current in the image channel. The third pulse is generated during the termination of the current pulse at the upper end of the antenna and the fourth one is produced by the termination of the current in the image of the channel. The reason why the propagation of the current pulse along the antenna does not produce any radiation is that the charge that gives rise to it moves without acceleration, since the current travels along the channel undistorted and at a constant speed. Observe that at any given point P, the radiation bursts generated by the initiation of the real current and the image current arrive simultaneously because they are generated at the same time and at the same point in space. In the analysis presented in the previous publications referenced earlier, only the energy transported by the first and second pulse is considered. Here we will consider the complete radiation field. In this respect, the analysis to be presented here is complete. In writing down the expressions for the radiation fields, we will assume that the point of observation is located far from the antenna. Note, however, that in our case the two ends of the antenna radiate as point sources.

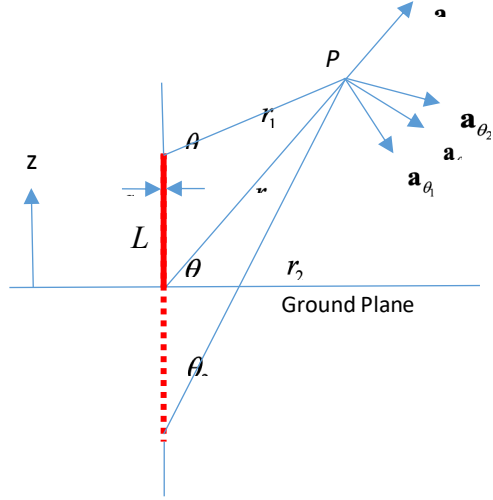

Figure 1: Geometry relevant to the problem under consideration. The perfectly conducting ground plane coincides with the x-y plane with z = 0. The image of the antenna in the ground plane is marked by the dotted line. The point of observation P and the relevant distances and angles are marked in the diagram. The unit vectors in the direction of increasing polar angles $\theta_1$, $\theta$ and $\theta_2$ are denoted by $\mathbf{a}_{\theta_1}$, $\mathbf{a}_\theta$ and $\mathbf{a}_{\theta_2}$ respectively. The length of the antenna is $L$ and its radius $a$.

The radiation field generated at the initiation of the current from the ground plane is given by [5,6]

$$\mathbf{E}_1(\theta,t) = \frac{i(t-r/c)v\sin\theta}{4\pi\varepsilon_0 c^2 r}\left\{\frac{1}{\left[1-\frac{v}{c}\cos\theta\right]}\right\}\mathbf{a}_\theta \qquad (1)$$

The radiation field resulting from the initiation of the current in the image of the antenna in the ground plane is given by

$$\mathbf{E}_2(\theta,t) = \frac{i(t-r/c)v\sin\theta}{4\pi\varepsilon_0 c^2 r}\left\{\frac{1}{\left[1+\frac{v}{c}\cos\theta\right]}\right\}\mathbf{a}_\theta \qquad (2)$$

The radiation field generated during the termination of the current at the upper end of the antenna is given by

$$\mathbf{E}_3(\theta_1,t) = -\frac{i(t-L/v-r_1/c)v\sin\theta_1}{4\pi\varepsilon_0 c^2 r_1}\left\{\frac{1}{\left[1-\frac{v}{c}\cos\theta_1\right]}\right\}\mathbf{a}_{\theta_1} \qquad (3)$$

Finally, the radiation field generated by the termination of the current in the image is given by

$$\mathbf{E}_4(\theta_2,t) = -\frac{i(t-L/v-r_2/c)v\sin\theta_2}{4\pi\varepsilon_0 c^2 r_2}\left\{\frac{1}{\left[1+\frac{v}{c}\cos\theta_2\right]}\right\}\mathbf{a}_{\theta_2} \qquad (4)$$

In the analysis to be presented in this paper, we will consider a current pulse propagating with the speed of light, $c$. In this case the radiation fields reduce to

$$\mathbf{E}_1(\theta,t) = \frac{i(t-r/c)\sin\theta}{4\pi\varepsilon_0 cr}\left\{\frac{1}{[1-\cos\theta]}\right\}\mathbf{a}_\theta \qquad (5)$$

$$\mathbf{E}_2(\theta,t) = \frac{i(t-r/c)\sin\theta}{4\pi\varepsilon_0 cr}\left\{\frac{1}{[1+\cos\theta]}\right\}\mathbf{a}_\theta \qquad (6)$$

$$\mathbf{E}_3(\theta_1,t) = -\frac{i(t-L/c-r_1/c)\sin\theta_1}{4\pi\varepsilon_0 cr_1}\left\{\frac{1}{[1-\cos\theta_1]}\right\}\mathbf{a}_{\theta_1} \qquad (7)$$

$$\mathbf{E}_4(\theta_2,t) = -\frac{i(t-L/c-r_2/c)\sin\theta_2}{4\pi\varepsilon_0 cr_2}\left\{\frac{1}{[1+\cos\theta_2]}\right\}\mathbf{a}_{\theta_2} \qquad (8)$$

Since the distance to the point of observation is much larger than the length of the antenna, we can make the following simplifications: $\cos\theta_1 \approx \cos\theta$, $\cos\theta_2 \approx \cos\theta$, $r \approx r_1 + L\cos\theta$ and $r \approx r_2 - L\cos\theta$. Moreover, the unit vectors $\mathbf{a}_{\theta_1}$ and $\mathbf{a}_{\theta_2}$ can be replaced by $\mathbf{a}_\theta$ and the distances $r_1$ and $r_2$ that appear in the denominator of these equations can be replaced by $r$. With these simplifications and combining the radiation fields given by equations (5) and (6) together, we obtain

$$E_{1,2}(\theta,t) = \frac{i(t-r/c)}{2\pi\varepsilon_0 cr}\left\{\frac{1}{\sin\theta}\right\}\mathbf{a}_\theta \qquad (9)$$

$$E_3(\theta,t) = -\frac{i(t-L/c\{1-\cos\theta\}-r/c)\sin\theta}{4\pi\varepsilon_0 cr}\left\{\frac{1}{[1-\cos\theta]}\right\}\mathbf{a}_\theta \qquad (10)$$

$$E_4(\theta,t) = -\frac{i(t-L/c\{1+\cos\theta\}-r/c)\sin\theta}{4\pi\varepsilon_0 cr}\left\{\frac{1}{[1+\cos\theta]}\right\}\mathbf{a}_\theta \qquad (11)$$

It is important to point out that these equations are valid in general irrespective of the duration of the current waveform and the length of the channel. They are also valid for the transmission line model of the lightning return strokes. One can observe that depending on the location of the point of observation and the duration of the current, these pulses may appear separately or they may overlap with each other. Let $\tau$ be the duration of the current pulse. One can see that the radiation pulses will start overlapping for locations of point P where $\theta < \theta_0$ with $\cos\theta_0 = 1 - \tau c / L$. Of course, we are assuming that $\tau < L/c$. In our analysis, we will consider current pulse durations which are much shorter than the time $L/c$. In this case one can show that almost all the energy transported by the radiation field is confined to the region where $\theta > \theta_0$ and the energy dissipated within the region $0 < \theta < \theta_0$ is negligible. Moreover, in this case, $\cos\theta_0$ becomes almost equal to unity and the value of $\theta_0$ reduces to $\theta_0 = \sqrt{2\tau c / L}$. The next step is to estimate the total energy transported by the radiation fields.

## 2.2 The energy transported by the radiation fields generated by a long antenna

Since the radiation field pulses occur without overlapping each other, the total energy transported by the radiation field is the sum of the energy transported by the four radiation field pulses. Consider the burst of radiation generated by the first two pulses. The Poynting vector associated with this radiation burst is given by

$$\mathbf{S}_{1,2}(\theta,t) = \frac{i(t-r/c)^2}{4\pi^2 \varepsilon_0 c r^2} \left\{ \frac{1}{\sin^2\theta} \right\} \mathbf{a}_r \quad (12)$$

The total Power transported by the pulse is given by

$$P_{1,2}(t) = \frac{i(t-r/c)^2}{4\pi^2 \varepsilon_0 c} 2\pi \int_{\theta_0}^{\pi/2} \left\{ \frac{1}{\sin\theta} \right\} d\theta \quad (13)$$

In writing down Equation (13), we are assuming that the power radiated for angles below $\theta_0$ is equal to zero. After solving the integral and noting that $\theta_0 = \sqrt{2\tau c / L}$, the total energy transported by the radiation bursts 1 and 2 is given by

$$U_{1,2} = \frac{1}{4\pi\varepsilon_0 c} \log\left[\frac{2L}{\tau c}\right] \int_{-\infty}^{\infty} i(t)^2 dt \quad (14)$$

Similarly, one can show that the energies transported by the radiation bursts $E_3$ and $E_4$ are given by

$$U_3 = \frac{1}{4\pi\varepsilon_0 c} \left\{ \log\left[\frac{L}{\tau c}\right] - \frac{1}{2} \right\} \int_{-\infty}^{\infty} i(t)^2 dt \quad (15)$$

$$U_4 = \frac{1}{4\pi\varepsilon_0 c} \left\{ \log 2 - \frac{1}{2} \right\} \int_{-\infty}^{\infty} i(t)^2 dt \quad (16)$$

Thus, the total energy transported by the radiation field is

$$U = U_{1,2} + U_3 + U_4 \quad (17)$$

Combining all the terms we obtain

$$U = \frac{\int_{-\infty}^{\infty} i(t)^2 dt}{2\pi\varepsilon_0 c} \left\{ \log\left[\frac{L}{\tau c}\right] - \frac{1}{2} \right\} \quad (18)$$

In the previous analysis reported in [3], a Gaussian current pulse was used to represent the current. One question related to that analysis is whether the results obtained are generally valid also for other current waveforms. To provide an answer to this here, we will consider four different families of current waveforms. They are the Power, Gaussian form, Sine form and Sinc form (truncated and modified) functions. They can be represented mathematically as:

Power:

$$i(t) = i_0 (1 - \frac{|t|^\alpha}{\delta^\alpha}) \qquad \alpha \geq 1 \text{ and } |t| \leq \delta \quad (19)$$

Gaussian form ($\alpha = 2$ gives the standard Gaussian function):

$$i = i_0 e^{-|t|^\alpha / \delta^\alpha} \qquad \alpha \geq 2 \quad (20)$$

Sine form ($\alpha = 1$ gives the standard sine function):

$$i = i_0 \sin^\alpha(\frac{\pi t}{\delta}) \qquad \alpha \geq 1 \qquad 0 \leq t \leq \delta \quad (21)$$

Sinc form: ($\alpha = 1$ gives the standard Sinc function):

$$i = i_0 \left(\frac{\sin t}{t}\right)^\alpha \qquad \alpha \geq 1 \qquad -\pi/2 \leq t \leq \pi/2 \quad (22)$$

In the above equations, $t$ is the time, $\delta$ is a positive constant and $i_0$ is the peak current. We define the duration of the current pulse $\tau$ as

$$\tau = \frac{\int_{-\infty}^{+\infty} i(t) dt}{i_0} \quad (23)$$

In the analysis to follow, we will use the Power family of current waveforms as an example, but the same procedure can be used in analyzing other families of current waveforms. With the above definition, the duration of the current pulse, $\tau$, pertinent to the Power family of current waveforms is given by

$$\tau = \frac{2\delta\alpha}{\alpha + 1} \quad (24)$$

The charge associated with this current pulse is given by

$$q = i_0 \tau \quad (25)$$

Thus, the expression for the current waveform can also be given in terms of this charge by

$$i(t) = \frac{q}{\tau}(1 - \frac{|t|^\alpha}{\delta^\alpha}) \qquad (26)$$

Substituting this into Equation (18) and performing the integration, we obtain the total energy transported by the radiation field as

$$U = \frac{q^2 \delta}{\tau^2 \pi \varepsilon_0 c}\left[1 - \frac{2}{\alpha+1} + \frac{1}{2\alpha+1}\right]\left\{\log\left(\frac{L}{\tau c}\right) - \frac{1}{2}\right\} \qquad (27)$$

Substituting for $\tau$ from Equation (24), we obtain

$$U = \frac{q^2(\alpha+1)^2}{4\delta\alpha^2 \pi \varepsilon_0 c}\left[1 - \frac{2}{\alpha+1} + \frac{1}{2\alpha+1}\right]\left\{\log\left(\frac{L}{\tau c}\right) - \frac{1}{2}\right\} \qquad (28)$$

Equation (28) gives the total energy transported by the radiation field generated by the current pulse as a function of the charge associated with the current pulse. During the generation of the radiation field, the charges in the current pulse either accelerate (during current initiation) or decelerate (during current termination) over the duration of the current pulse which is equal to $\tau$. This indeed is the duration over which part of the energy of the system is transformed into the radiation field. Let us define the action $A$ associated with the radiation field as the product of the total transported energy and this duration. Thus, the action associated with the radiation field is given by

$$A = \frac{q^2}{2\pi\varepsilon_0 c}\left[\frac{(\alpha+1)}{\alpha} - \frac{2}{\alpha} + \frac{\alpha+1}{\alpha(2\alpha+1)}\right]\left\{\log\left(\frac{L}{\tau c}\right) - \frac{1}{2}\right\} \qquad (29)$$

Note that the action has the units of angular momentum.

## 2.3 Absolute maximum value of the action associated with the radiation field and its consequences

Let us estimate the absolute theoretical maximum value of the action ever possible associated with the radiation field. For a given charge, the action increases with increasing $L/\tau c$. It is important to point out here that the radiation field as given earlier is valid when $\tau c$ is on the order of the radius of the conductor or larger. This is the case because if this condition is not satisfied, the radiation fields generated from different parts of the cross section of the antenna may interfere destructively reducing the amplitude of the radiation field. Thus, the smallest value of $\tau c$ for which the equations are valid is obtained when $\tau c \approx a$, where $a$ is the radius of the conductor. With this change, the maximum action for a given charge is obtained for the largest possible value of the ratio $L/a$. Let us consider the natural limits imposed on this ratio by nature. The smallest possible radius of an antenna that can exist in nature is equal to the Bohr radius $a_0$, i.e., the atomic dimensions. The largest possible value of the antenna length that one can have in nature is equal to the radius of the universe. Since the universe is expanding, the radius of the universe changes with time. However, according to the current understanding, the radius of the universe (the region where events are in causal contact) becomes constant at some future epoch and this value is given by the steady state value of the Hubble radius [7, 8, 9, 10]. Let us denote this radius by $R_\infty$. Substituting this into our expression we obtain

$$\langle A \rangle_{max} = \frac{q^2}{2\pi\varepsilon_0 c}\left[\frac{(\alpha+1)}{\alpha} - \frac{2}{\alpha} + \frac{\alpha+1}{\alpha(2\alpha+1)}\right]\left\{\log\left(\frac{R_\infty}{a_0}\right) - \frac{1}{2}\right\} \quad (30)$$

In Equation (30), $\langle A \rangle_{max}$ is the absolute maximum value of the action that can be achieved by a given charge. Based on general relativity, the steady state value of the Hubble radius is given by $R_\infty = c^2\sqrt{3/8\pi G\rho_\Lambda}$ where G is the gravitational constant and $\rho_\Lambda$ is the dark energy density [7]. Substituting this expression into Equation (30) yields

$$\langle A \rangle_{max} = \frac{q^2}{2\pi\varepsilon_0 c}\left[\frac{(\alpha+1)}{\alpha} - \frac{2}{\alpha} + \frac{\alpha+1}{\alpha(2\alpha+1)}\right]\left\{\log\left(\frac{c^2\sqrt{3/8\pi G\rho_\Lambda}}{a_0}\right) - \frac{1}{2}\right\} \quad (31)$$

An expression similar to that given by Equation (31) can be obtained for other families of current waveforms by following the same procedure. Let us now study how the value of this action changes as a function of the charge associated with different families of current waveforms.

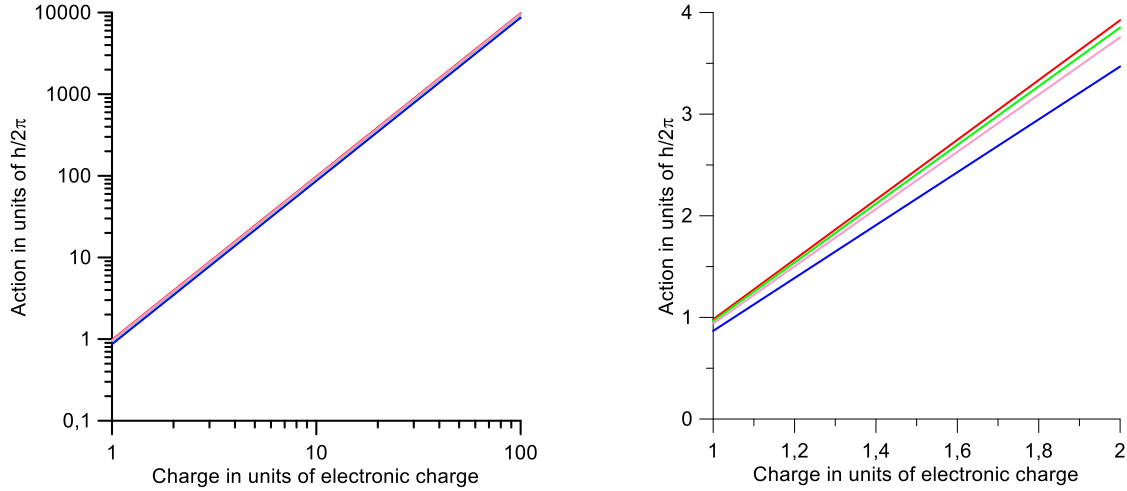

(a)          (b)

Figure 2: (a) The absolute maximum value of the action $\langle A \rangle_{max}$ as a function of the charge for different current waveshapes. (1) Power with $\alpha = 2$ (red). (2) Gaussian with $\alpha = 2$ (blue). (3) Sine with $\alpha = 1$ (green). (4) Sinc with $\alpha = 1$ (pink). (b) The same but on an expanded and linear scale.

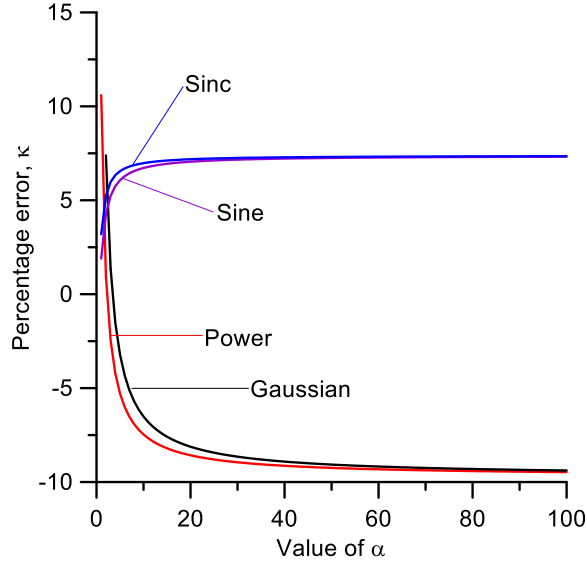

Figure 3: The percentage difference between the value of $\langle A \rangle_{max}$ and $h/2\pi$ when $q = e$ as a function of $\alpha$ for different current waveshapes.

Figure 2 shows a plot of $\langle A \rangle_{max}$ as a function of the magnitude of the charge for the four families of current waveforms for several values of $\alpha$. We have represented the action in units of $h/2\pi$, which is the atomic unit of the angular momentum with $h$ representing the Planck constant. According to the results presented in Figure 2, Equation (31) and the corresponding ones pertinent to other families of current waveforms predict that when $q = e$, where $e$ is the elementary charge, $\langle A \rangle_{max} \approx h/2\pi$. In order to investigate how this relationship will change for different values of $\alpha$, let us define the parameter $\kappa$ as the percentage difference between the value of $\langle A \rangle_{max}$ and $h/2\pi$ when $q = e$ for a given value of $\alpha$. The variation of $\kappa$ as a function of $\alpha$ is plotted in Figure 3. Note that over a broad range of the values of $\alpha$ (1 to 100), the value of $\langle A \rangle_{max}$ differs only by less than 10% of $h/2\pi$ when $q = e$. Since $\langle A \rangle_{max}$ is the absolute maximum action that can be realized by any given antenna, the results can be summarized by the mathematical statement $A \geq h/2\pi \Rightarrow q \geq e$ where $A$ is the action of the radiated energy of the antenna and $q$ is the magnitude of the charge associated with the current. Since we have considered the maximum ratio $L/a$ ever possible by an antenna, this mathematical statement is universal and it is satisfied by any arbitrary wire antenna. Note that in the previous work [1, 2], $h/4\pi$ instead of $h/2\pi$ appears in the mathematical statement. This difference in the factor of 2 was caused by the fact that the work done in previous publications did not take into account all the radiation field pulses generated by the antenna in estimating the energy transported by the radiation field. It is important to point out here that the reverse of this mathematical statement, i.e., $q \geq e \Rightarrow A \geq h/2\pi$ is not valid. This is the case because even in the cases where the charge is larger than the electronic charge, by decreasing the size of the antenna one can satisfy the condition $A < h/2\pi$.

## 3. Frequency domain electromagnetic fields

### 3.1 Radiation fields and the transport of energy

As the radiating system, we will consider a monopole antenna of length $L$ located over a perfectly conducting ground plane. The relevant geometry is similar to that of Figure 1. As before, the perfectly conducting plane coincides with the x-y plane with z = 0. The antenna is located along the z-axis. The antenna is fed by a sinusoidal current from the ground end (i.e., z = 0). If one neglects the dissipation losses, it is a reasonable approximation to assume that the current distribution along the antenna is given by [11, 12, 13]

$$I(t,z) = I_0 \sin\left\{\frac{2\pi}{\lambda}(L-z)\right\} e^{j\omega t} \quad 0 \le z \le L \tag{32}$$

The current distribution in the image of the antenna in the ground plane is given by

$$I(t,z) = I_0 \sin\left\{\frac{2\pi}{\lambda}(L+z)\right\} e^{j\omega t} \quad -L \le z \le 0 \tag{33}$$

In references [11, 12], one can find the conditions under which this approximation is valid. In general, it is valid if one can neglect all the losses including thermal, dispersive, and dissipative losses.

The electric and magnetic fields at large distances generated by this antenna are given by [11]

$$E_\theta = \frac{j\eta I_0 e^{j\omega t} e^{-jkr}}{2\pi r} \left[\frac{\cos(kL\cos\theta) - \cos(kL)}{\sin\theta}\right] \tag{34}$$

$$H_\phi = \frac{j I_0 e^{j\omega t} e^{-jkr}}{2\pi r} \left[\frac{\cos(kL\cos\theta) - \cos(kL)}{\sin\theta}\right] \tag{35}$$

where $k = 2\pi/\lambda$ and η is the impedance of free space. The average (over a period of oscillation) Poynting vector associated with these fields is given by

$$\langle \mathbf{S} \rangle = \frac{1}{2}\mathrm{Re}\left[\mathbf{E} \times \mathbf{H}^*\right] \tag{36}$$

From this, (35), and (36), we obtain

$$\langle \mathbf{S} \rangle = \frac{\eta I_0^2}{8\pi^2 r^2}\left[\frac{\cos(kL\cos\theta) - \cos(kL)}{\sin\theta}\right]^2 \mathbf{a_r} \tag{37}$$

The total average power dissipated over a single period of oscillation by the antenna is then given by

$$P_{av} = \frac{I_0^2}{4\pi\varepsilon_0 c} \int_0^{\pi/2} \left[ \frac{\cos(kL\cos\theta) - \cos(kL)}{\sin\theta} \right]^2 \sin\theta d\theta \quad (38)$$

If the oscillating charge associated with the current is given by $Q = qe^{j2\pi vt}$, the above equation can be written as

$$P_{av} = \frac{q^2(2\pi v)^2}{4\pi\varepsilon_0 c} \int_0^{\pi/2} \left[ \frac{\cos(kL\cos\theta) - \cos(kL)}{\sin\theta} \right]^2 \sin\theta d\theta \quad (39)$$

The integral in (39) can be solved analytically (see reference [10]) resulting in the following expression

$$P_{av} = \frac{q^2(2\pi v)^2}{8\pi\varepsilon_0 c} \left\{ \begin{array}{l} \gamma + \ln(2kL) - C_i(2kL) + \frac{1}{2}\sin(2kL)\left[S_i(4kL) - 2S_i(2kL)\right] \\ + \frac{1}{2}\cos(2kL)\left[\gamma + \ln(kL) + C_i(4kL) - 2C_i(2kL)\right] \end{array} \right\} \quad (40)$$

where $C_i$ is the cosine integral, $S_i$ is the sine integral and $\gamma$ is Euler's constant.

Observe first that the power generated by the antenna consists of bursts of energy of duration $T/2$ where $T$ is the period of oscillation. The energy dissipated by a single burst of energy $U$ of duration $T/2$, is given by

$$U = \frac{q^2\pi v}{4\varepsilon_0 c} \left\{ \begin{array}{l} \gamma + \ln(2kL) - C_i(2kL) + \frac{1}{2}\sin(2kL)\left[S_i(4kL) - 2S_i(2kL)\right] \\ + \frac{1}{2}\cos(2kL)\left[\gamma + \ln(kL) + C_i(4kL) - 2C_i(2kL)\right] \end{array} \right\} \quad (41)$$

One can observe from this equation that for large values of $kL$, the value of $U$ oscillates rapidly with $kL$. The upper and lower bounds of $U$ occur when $2kL = n\pi$ and $2kL = m\pi$ where $n$ and $m$ are even and odd integers (i.e., when $\cos(kL) = 1$ or $\cos(kL) = -1$). The median value of $U$ is given by

$$U_{med} = \frac{q^2\pi v}{4\varepsilon_0 c} \{\gamma + \ln(2kL) - C_i(2kL)\} \quad (42)$$

Observe that for large values of $kL$, the cosine integral varies as $\cos(2kL)/(2kL)^2$ and it can be neglected with respect to other terms. Thus, for large values of $kL$, the expression for the median energy reduces to

$$U_{med} = \frac{q^2\pi v}{4\varepsilon_0 c} \{\gamma + \ln(2kL)\} \quad (43)$$

### 3.2 Absolute maximum value of the median energy as a function of the oscillating charge

The equations derived for the radiation fields of the frequency domain antenna are valid provided that $\lambda \geq a$, where $a$ is the radius of the antenna. Using this fact, one can see from Equation (43) that for a given antenna length, the maximum value of energy is reached for wavelengths comparable to the radius of the antenna $a$. Using the same arguments as in the case of the time domain antenna, in order to get the absolute maximum value of the energy transported by the radiation fields, we will replace $a$ by the Bohr radius and $L$ by the radius of the universe. Thus, the upper limit of the median energy dissipated within a single power burst of duration $T/2$ for a given charge is given by,

$$\langle U_{med} \rangle_{max} = \frac{q^2 \pi \nu}{4\varepsilon_0 c} \{\gamma + \ln(4\pi R_\infty / a_0)\} \quad (44)$$

Figure 4 shows a plot of $\langle U_{med} \rangle_{max}$ as a function of the magnitude of the oscillating charge. This equation predicts that when $q = e$, where $e$ is the elementary charge, $\langle U_{med} \rangle_{max} \approx h\nu$ where $h$ is the Planck constant. Since $\langle U_{med} \rangle_{max}$ is the maximum energy that can be radiated within a single burst of power for any given charge, the results can be summarized by the mathematical statement

$U \geq h\nu \Rightarrow q \geq e$ where $U$ is the energy radiated in a single power burst of duration $T/2$, $h$ is the Planck constant, $\nu$ is the frequency of oscillation and $q$ is the peak magnitude of the oscillating charge associated with the current. Since we have considered the maximum ratio $L/a$ ever possible by an antenna, this mathematical statement is universal and it is satisfied by any arbitrary antenna. Moreover, as in the case of the time domain antenna, it is important to point out that the reverse of this mathematical statement, i.e., $q \geq e \Rightarrow U \geq h\nu$ is not necessarily true. For example, even if the charge is larger than the electronic charge, by decreasing the size of the antenna one can make the energy be sufficiently low for the relation $U < h\nu$ to be satisfied.

Furthermore, the derivation is purely based on classical electrodynamics and the appearance of the Planck constant in the expression is due to the use of atomic units to describe the energy.

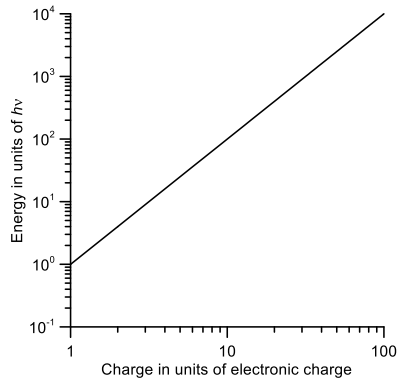

Figure 4: The absolute value of the median energy transported by a single burst of radiation, $\langle U_m \rangle_{max}$, as a function of the charge $q$ as given by Equation (44).

## 4. Discussion and the significance of the results

Let us consider the results presented in Figure 3. Observe that in the case of Power and Gaussian current waveforms, there is a particular value of $\alpha$ that makes $\langle A \rangle_{max} = h/2\pi$ when $q = e$ (i.e., $\kappa = 0$). Even though the value of $\alpha$ that satisfies this condition is different in the Power and Gaussian waveforms, our analysis shows that the current waveshapes corresponding to those values of $\alpha$ are almost identical. Even in the other families, the current waveforms corresponding to the value of $\alpha$ that gives the smallest value of $\kappa$ are also similar to the latter current waveshape. In the case of the Power family of current waveforms given by Equation (19), the value of $\alpha$ that makes $\langle A \rangle_{max} = h/2\pi$ when $q = e$ is equal to about 2.2 (when $\alpha = 2$, $\langle A \rangle_{max}$ differs by about 1% from $h/2\pi$).

It is interesting to point out that the current waveshapes as given by the power family of current waveforms can be related to the currents associated with the motion of different charge distributions. For example, consider the case of a moving sphere with a uniform charge density. In this case, the current associated with the charge distribution is given by Equation (19) with $\alpha = 2$. If, on the other hand, the charge density in the sphere varies linearly with the radius (i.e., it is maximum at the surface and zero at the center), the corresponding current is obtained by replacing $\alpha = 3$ in the Power family of current waveforms. As the value of $\alpha$ increases, the charge will become more and more concentrated close to the surface of the sphere.

Note that the ultimate antenna that we have considered here is only a theoretical possibility and it could never be realized in practice. Moreover, in deriving the results presented here we have assumed ideal conditions without losses, which also cannot be realized in practice. In this respect, we have to treat the problem we have analyzed as a 'Gedanken' experiment. The goal of such an experiment here is to extract the ultimate limits of a given theory.

One may question whether the mathematical statement derived here (i.e., $A \geq h/2\pi \Rightarrow q \geq e$) will be disturbed by the losses that will always be present in the actual antenna systems. The effect of losses is to reduce the radiated energy associated with a given current waveform (or charge) and, for this reason, the derived mathematical statement will still be valid even in the presence of losses. It is important to point out that the derived mathematical statement has to be interpreted as an order of magnitude estimation. For example, the value of the charge that makes the action equal to $h/2\pi$ may change slightly from one type of current waveform to another, but our analysis shows that the charge always remains within the order of magnitude of the electronic charge. Moreover, it depends to some extent on the definition of the duration of the current waveform. Observe also that we have obtained this result by setting the ratio of the antenna length to the antenna radius by the ratio of the size of the universe to the atomic dimensions. However, this ratio appears inside a logarithmic term and the dependence of the estimated action on this ratio is rather weak.

The discussion concerning the effects of losses given above for the time domain antenna is also applicable to the results obtained for the frequency domain antenna.

Observe that the presence of Planck constant in the inequalities derived here might incorrectly give the impression that quantum mechanics is involved in this derivation. This is not the case. The derivation is purely classical without any involvement of quantum mechanics. It is only our choice of units for the angular momentum and energy that introduced the Planck constant into the equations. But the most interesting observation one can make is that just by knowing the experimental fact that the minimum charge that can be found in nature is the electronic charge, one could have heralded the emergence of quantum mechanics by using purely the electromagnetic field equations of classical electrodynamics.

In this paper we have derived two inequalities, one for the time domain antenna and the other for the frequency domain. The left-hand side of the mathematical statements are $U\tau \geq h/2\pi$ for the time domain antenna and $U \geq h\nu$ for the frequency domain antenna. The validity of these two inequalities cannot be assertained within the confines of the classical electrodynamics. To do that we have to appeal to quantum mechanics. First, let us assertain the general validity of these inequalities and, based on that, discuss the significance of the results.

As mentioned earlier, the derivation presented in this paper is based purely on equations based on classical electrodynamics and general relativity. Moreover, the description of the current even when the charge associated with it is reduced to the electronic charge is also based on classical physics. This classical analysis led to two mathematical statements, the significant of which will be discussed in the following sections.

Consider the right-hand portion of the mathematical statements obtained in the paper. The right-hand portion of the derived mathematical statements state that $q \geq e$. This is actually not a consequence of classical electrodynamics because the theory does not specify any limits on the magnitude of the charge. However, from experimental observations, we know today that the minimum free charge that exists in nature is equal to the electronic charge or the elementary charge. This shows that the two inequalities $U \geq h\nu$ and $U\tau \geq h/2\pi$ appearing in the two mathematical statements have to be valid. Let us consider in what follows these two inequalities.

First consider the inequality $U\tau \geq h/2\pi$. Let us assume that this condition is valid and look for the consequences. This condition indicates that for any given value of $\tau$ over which the system radiates away its energy, the energy dissipated cannot be smaller than the value $h/2\pi\tau$. In other words, there is a lower limit to the value of radiated energy that can be measured for a given duration of the emission. This also indicates that in any energy measurement, the error associated cannot be smaller than a value $\Delta U$ given by the equation

$$\Delta U \approx h/2\pi\tau \quad (45)$$

The question is whether the above conclusion is valid. To find an answer to this question, let us consider quantum mechanics. The problem could be visualized as follows. The antenna radiates over a time interval given by $\tau$. During this time, a large number of photons with different energies are emitted. If we consider any of these photons, we are unable to predict when that photon will be emitted except that it will be emitted during a time interval $\tau$. According to Heisenberg's time-energy uncertainty principle, this will introduce an uncertainty in the energy of the photon equal to $h/2\pi\tau$. This is true for any given photon irrespective of its energy. This in turn leads to a minimum uncertainty in the total energy of the radiation, $\Delta U$ equivalent to

$\Delta U \approx h/2\pi\tau$ (46)

The two relationships given by equations (45) and (46) are identical. Thus, one can utilize quantum mechanics to argue that the condition $A \geq h/2\pi$ is valid. Even though the mathematical statement $A \geq h/2\pi \Rightarrow q \geq e$ is derived purely based on classical electrodynamics, there is a quantum mechanical justification for the left hand side of this mathematical statement. This shows that the left-hand side of the mathematical statement is dictated by the quantum mechanics and the condition q ≥ e is a consequence of quantum mechanics.

Let us now consider the inequality $U \geq h\nu$. Recall that $U$ is the energy dissipated within the time $T/2$, the duration of a single burst of radiation. According to quantum mechanical interpretation, the electromagnetic radiation consists of photons. If any electromagnetic energy is dissipated within the time duration $T/2$, then at least one or more photons should be released by the radiator during this time interval. Since the energy of a photon cannot be smaller than $h\nu$, the energy released during the time period $T/2$ cannot be smaller than $h\nu$. In other words, the quantum mechanical nature of electromagnetic radiation gives rise to the inequality $U \geq h\nu$.

This discussion shows that not only the right-hand portion of the mathematical statements but also the left-hand portion of the two mathematical statements given below are correct.

$A \geq h/2\pi \Rightarrow q \geq e$ (47)

$U \geq h\nu \Rightarrow q \geq e$ (48)

This shows that the condition $q \geq e$ is a consequence of the quantum nature of the electromagnetic fields. The results are significant in two ways. First, knowing the fact that the electronic charge is the minimum charge, the classical electrodynamics itself could have come up with the two most fundamental equations in quantum mechanics: the quantum nature of electromagnetic radiation and the uncertainty principle. Second, the fact that the minimum free charge that exists in nature is equal to the electronic charge is shown for the first time to be a direct consequence of the photonic nature of the electromagnetic fields.

Consider the expression for the energy we have obtained in the case of the frequency domain antenna (here we consider the results from frequency domain because $U \geq h\nu$ is exact whereas, as discussed earlier, $A \geq h/2\pi$ is an order of magnitude estimation). Now, based on experimental investigations $q \geq e$ and based on quantum nature of the electromagnetic fields, the minimum energy that can be radiated within a single burst of radiation has to be larger than or equal to $h\nu$. Thus, we can write

$$\frac{q^2 \pi \nu}{4\varepsilon_0 c}\{\gamma + \ln(4\pi R_\infty / a_0)\} \geq h\nu \Rightarrow q \geq e \quad (49)$$

Considering the equal signs of inequalities on both sides we obtain the expression

$$\frac{e^2 \pi}{4\varepsilon_0 c} \{\gamma + \ln(4\pi R_\infty / a_0)\} = h \qquad (50)$$

In the above equation, $R_\infty$ is the steady-state value of the Hubble radius. Observing that $R_\infty = c^2 \sqrt{3/8\pi G \langle \rho_\Lambda \rangle}$ where $\langle \rho_\Lambda \rangle$ is the dark energy density [7], the above equation can be written as

$$\frac{e^2 \pi}{4\varepsilon_0 c} \{\gamma + \ln(4\pi c^2 \sqrt{3/8\pi G \langle \rho_\Lambda \rangle} / a_0)\} = h \qquad (51)$$

This equation connects the electronic charge to the dark energy density through other constants of nature. By making the dark energy density the subject, we can write

$$\langle \rho_\Lambda \rangle = \left\{ \frac{24\pi^3 m_e^2 c^6 e^{2\gamma}}{Gh^2} \Gamma^2 \right\} e^{-\frac{4}{\pi \Gamma}} \qquad (52)$$

Note that we have replaced $a_0$ by $h/2\pi m_e c \Gamma$ where $m_e$ is the rest mass of the electron and $\Gamma$ is the fine structure constant which is equal to $e^2/2\varepsilon_0 ch$. This equation predicts $\langle \rho_\Lambda \rangle$ = 4.3x10$^{-10}$ J/m$^3$ which is in agreement with the current experimentally observed value of 5.356x10$^{-10}$ J/m$^3$ [8, 9, 10]. This is a significant result given the current state of the theoretical estimations of dark energy density. This equation, which combines together the dark energy, electronic charge and mass, speed of light, gravitational constant and Planck constant, creates a link between classical field theories (i.e., classical electrodynamics and general relativity) and quantum mechanics.

After the observation of the accelerated expansion of the universe, scientists have invoked dark energy as a possible explanation for this rapid expansion. However, several attempts to theoretically estimate the dark energy density gave rise to values which are about $10^{120}$ times larger than the observed value [14]. Modern research work led to a value which is about $10^{60}$ larger than the observed value. This discrepancy, known as the vacuum energy catastrophe, is named as one of the worst predictions in physics. The results presented here provide an equation that predicts the value of the dark energy density in terms of other fundamental constants to a reasonable accuracy.

Observe that the dark energy density is strongly connected to the fine-structure constant which determines the strength of the electromagnetic force. According to Equation (52), the dark energy density decreases with decreasing fine-structure constant and vice versa. To the best of our knowledge, this is the first time that a relationship is derived for the dark energy in terms of the other fundamental constants in nature based on classical field theory (general relativity and classical electrodynamics) in combination with the quantization of energy. The narrative presented here is somewhat similar to the estimation of the Bekenstein-Hawking entropy of a black hole based on general relativity and thermodynamics (which similarly is overestimated using standard quantum field theories) [15, 16].

It is interesting to note that based on the Large Number Hypothesis of Dirac [17] it had been speculated based on numerical coincidences that the fine-structure constant, $\Gamma$, changes as the inverse of the logarithm of the age of the universe expressed in Planck units [18]. Consider Equation (50). It can be written as

$$\Gamma \approx \frac{1}{\left\{\ln(4\pi \langle R_\infty \rangle / a_0)^{\pi/2}\right\}} \quad (53)$$

After substituting for the parameters inside the logarithmic term we obtain

$$\Gamma = \frac{1}{\ln(1.326 \times 10^{59})} \quad (54)$$

The number inside the logarithmic term in the above equation is approximately equal to the current age of the universe in Planck units of time. This, therefore, provides a possible explanation for the numerical coincidence $\Gamma \sim 1/\ln(t)$ where $t$ is the current age of the universe in Planck units. This point illustrates that extracting information from numerical coincidences in nature should be done with caution.

While some physicists are trying to understand and resolve the vacuum energy catastrophe, others are suggesting that, similar to the electronic charge or the fine-structure constant, one should treat the dark energy density as a natural constant whose value is fixed by experimental observations. Some physicists raise the philosophical question: What is the reason for the dark energy density to have an extremely small value instead of zero? Equation (52) provides an answer to this question. This equation shows that the magnitude of the dark energy density is decided by the value of the fine-structure constant. In this respect dark energy density could be treated as a dependent parameter. The equation also shows that a zero dark energy density will arise only in a universe which does not contain the electromagnetic force. It is also of interest to note that Equation (52) consists of two factors: One factor (the one inside the curly bracket) is 75 orders of magnitude larger than the experimentally measured value of the dark energy density but the second factor delicately balances this large number to reduce the dark energy density to the experimentally observed value. We hope the latter will provide some insights to the physicists who attempt to calculate the dark energy density using standard procedures.

## 5. Conclusion

The minimum energy that can be generated by an electromagnetic radiator with a length equal to the ultimate radius of the universe is analyzed. From this analysis, we have obtained an expression for the dark energy density (in effect Einstein's cosmological constant) using gravitational and electrodynamical constants and the quantum nature of energy. The physical connection we obtained between dark energy density and the fine-structure constant (or electronic charge) could be an indication that the nature of dark energy, at least to a large extent, is electromagnetic. Moreover, our study shows that (i) knowing the fact that the electronic charge is the minimum charge, the classical electrodynamics itself could have come up with the two most fundamental equations in quantum mechanics: the quantum nature of electromagnetic radiation and the uncertainty principle; and (ii) the fact that the minimum free charge that exists in nature is equal to the electronic charge is shown for the first time to be a direct consequence of the photonic nature of the electromagnetic fields.

We believe that further work into these ideas, while providing more convincing arguments to the latter, could generate more insights into the nature of dark energy of the universe.

**Data Availability**: The datasets used and/or analysed during the current study available from the corresponding author on reasonable request.